**Comparing acoustic analyses of speech data collected remotely**

Cong Zhang,[1] Kathleen Jepson,[1] Georg Lohfink,[2] Amalia Arvaniti,[1]

[1] *Faculty of Arts, Radboud University, Nijmegen, GLD, 6500 HD, Netherlands*

[2] *School of European Culture and Languages, University of Kent, Canterbury, Kent, CT2 7NF, UK*

Face-to-face speech data collection has been next to impossible globally due to COVID-19 restrictions. To address this problem, simultaneous recordings of three repetitions of the cardinal vowels were made using a Zoom H6 Handy Recorder with external microphone (henceforth *H6*) and compared with two alternatives accessible to potential participants at home: the Zoom meeting application (henceforth *Zoom*) and two lossless mobile phone applications (Awesome Voice Recorder, and Recorder; henceforth *Phone*). F0 was tracked accurately by all devices; however, for formant analysis (F1, F2, F3) Phone performed better than Zoom, i.e. more similarly to H6, though data extraction method (VoiceSauce, Praat) also resulted in differences. In addition, Zoom recordings exhibited unexpected drops in intensity. The results suggest that lossless format phone recordings present a viable option for at least some phonetic studies.

## I. INTRODUCTION

Speech production studies have been significantly impacted by restrictions related to COVID-19, as both access to laboratories and face-to-face interaction with study participants have been restricted. In order to adapt to the situation, we set out to test whether alternatives easily accessible to participants recorded remotely can produce recordings suitable for acoustic analysis. We note that these findings are of interest to phoneticians working on speech production even if COVID-19 related restrictions are completely lifted in some countries, as researchers may continue to have limited access to speech communities in some countries. More generally, researchers may have to conduct recordings remotely for other reasons, for instance because of ethical, political, or financial restrictions that make travel difficult or impossible.

Research has already examined the performance of several devices that can be used for recordings, such as iPads (De Decker, 2016; Maryn et al., 2017), computers (De Decker and Nycz, 2011; Kojima et al., 2019; Vogel et al., 2014), and smart phones (Grillo et al., 2016; Kojima et al., 2019; Manfredi et al., 2017; Uloza et al., 2015; Vogel et al., 2014; see Jannetts et al., 2019, for a review). Other studies have examined the effects of different file formats, such as lossless Apple .m4a files (De Decker and Nycz, 2011), lossy compressed .mp3 files (Bulgin et al., 2010), and audio extracted from compressed video files (De Decker and Nycz, 2011).

Two key findings emerge from these studies. First, $F_0$ is often unaffected by recording device and file format (Bulgin et al., 2010; Fuchs and Maxwell, 2016; Jannetts et al., 2019; Maryn et al., 2017), though it is unclear whether this applies equally well to $F_0$ that exhibits significant dynamic changes, as it does to steady $F_0$ (which is what is typically tested). Second, lossy formats distort the F1-F2 vowel space in unpredictable ways; both expansion and compression (i.e., changes in both F1 and F2 simultaneously) are observed inconsistently across speakers (Bulgin et al., 2010), with women's speech showing greater distortion in lossy files recorded in quiet conditions (De Decker and Nycz, 2011). Noise, on the other hand, can lead to greater vowel space distortion in male voices instead (De Decker, 2016).

We add to this line of research, by comparing recordings made with a high-quality digital recorder, Zoom H6 Handy Recorder (henceforth *H6*) with recordings made using two "remote" options: the Zoom cloud meeting application (henceforth *Zoom*) and mobile phone applications that produce sound files in lossless formats (henceforth *Phone*). We investigated these two options because they are convenient, free, readily available, and allow for local file storage. Phone-based options have the benefit of only requiring a smartphone to use, which most people have ready access to. They are a convenient way to record lossless format files which are recorded by H6, and are used as standard in acoustic research. Zoom has been successfully used for supervised online data collection (Leemann et al., 2020) and so may be a convenient recording tool already in use in remote data collection. Zoom was also selected as 1) participants do not need to have a personal account to join a Zoom meeting, which may be relevant for data protection obligations; 2) the built-in recording function in Zoom allows local recording without relying on internet connection. It is noted that Zoom does require internet connection to start a meeting session.  Local recording, however, does not rely on the internet connection quality. Other computer-based options may need a paid subscription, are browser-based, or require fast, stable internet connection. These features pose two problems: use requires suitable infrastructure in the locations where the data are collected, and this may not always be available; further, data saved in proprietary applications could create issues with data storage and personal data protection regulations. Zoom recordings may be comparable to those made using other conferencing software such as Skype and Microsoft Teams (Freeman & De Decker, 2021);

However, further investigation is needed to generalise the findings of this study to different applications.

## II. METHODS

### A. Participants

Four females (PF1-4) and three males (PM1-3), aged 30-52 (mean 37) took part in the study. The number of participants was limited due to health safety concerns and COVID-19 restrictions at the time of recording in mid-2020; these are factors that have impacted remote data collection studies world-wide (see e.g., Freeman and De Decker, 2021, Sanker et al., 2021 who analysed the speech of two and three speakers respectively). PF1, PF2, PF4, and PM3 had specialised phonetics training, while PF3, PM1, and PM2 had broad training in linguistics. PF2, PF4, and PM2 were monolingual speakers of Australian English; the other four were multilingual with Mandarin (PF1), Bengali (PF3), Kurdish (PM1), and German (PM3) as L1s. The variable linguistic backgrounds of the participants are not a problem for the present study which focuses on differences between devices and thus on within-speaker comparisons. All participants were aware of the purpose of the study. Though this is a small sample size, it is hoped the results will be of help to speech researchers.

### B. Materials

The materials consisted of pure tones and elicitation of sustained versions of the primary cardinal vowels, [i, e, ɛ, a, ɑ, ɔ, o, u]. Here we report only on the results from the vowel recordings. We used sustained vowels to make our findings comparable to those of previous studies (e.g., Grillo et al., 2016; Manfredi et al., 2017; Maryn et al., 2017; Uloza et al., 2015;

Vogel et al., 2014), which in turn used sustained vowels because they "feature simple acoustic structure and allow reliable detection and computation of acoustic features" (Uloza et al., 2015).

### C. Recording devices and applications

Participants made simultaneous recordings of the vowels using an H6 with an external microphone, a Phone running a recording application using the built-in microphone, and a laptop running Zoom using the built-in microphone of the laptop; see Table I for details. The range of mobile phones and computers used simulates real-world scenarios where participants in a remote speech production study would use different devices.

TABLE I. Recording equipment and application information for each participant.

| Participant ID | Mobile phone | App | Zoom | Recorder | Microphone |
|---|---|---|---|---|---|
| **PF1** | Samsung Note10 | AVR[a] | MS Surface Pro 6 | Zoom H6 | Sennheiser HSP2 |
| **PF2** | Samsung Galaxy S10e | AVR | Dell Precision 5520 | Zoom H6 | Rode NT3 cardioid mic (on stand) |
| **PF3** | Samsung Note10 | AVR | MS Surface Pro 6 | Zoom H6 | Sennheiser HSP2 |
| **PF4** | Google Pixel 3a | AVR | Lenovo Thinkpad T495 | Zoom H6 | Sennheiser HSP2 |
| **PM1** | Samsung Note9 | AVR | MS Surface Pro 6 | Zoom H6 | Sennheiser HSP2 |
| **PM2** | Apple iPhone 5s | AVR | Lenovo Thinkpad T495 | Zoom H6 | Sennheiser HSP2 |
| **PM3** | bq AQUARIS E4.5 Ubuntu Edition | Recorder[b] | Lenovo Thinkpad T495 | Zoom H6 | Sennheiser ME64 & K6P |

[a] Awesome Voice Recorder (Newkline, 2020)

[b] Recorder (DawnDIY, 2016) is an app available on Linux phones; see section II D. for details.

### D. Recording procedure

All recordings were made in quiet home locations, using H6, Phone and Zoom simultaneously. The H6 recorded mono .wav files, at 44.1 kHz, 24 bits. Phone was used with two applications:

Awesome Voice Recorder (henceforth *AVR*; Newkline, 2020) available on Android and iOS phones, and Recorder (DawnDIY, 2016) available on Linux phones. Both recorded mono-channel files, at 44.1 kHz, 256 bps, in lossless formats (.wav for AVR and .ogg for Recorder). Zoom v. 5.1.2 (28642.0705) with default settings was used to record stereo-channel .m4a files, a lossy format; files were saved locally. We note that the "enable original sound" option, available from v. 5.2.2 (45108.0831) in September 2020, had not been released at the time of data collection. The .ogg and .m4a files were converted to mono-channel .wav files at 44.1 kHz and 256 bps using VLC (VideoLan, 2019).

Following previous studies (e.g., Grillo et al., 2016; Manfredi et al., 2017; Maryn et al., 2017; Uloza et al., 2015; Vogel et al., 2014), we instructed participants to produce and sustain each vowel for 3-5 s, and repeat them three times

The devices were placed as follows: the Zoom computer was placed on a table directly in front of the participant, approximately 40-50 cm away, resembling a Zoom meeting set up; the participant held the Phone approximately 10-20 cm from their mouth; the H6 was used with either a head-mounted microphone or a microphone with pop filter on a stand 15 cm in front of the participant. Participants were asked to turn all devices to silent mode. Participants were not asked to restart their devices, or turn off all other processes and background applications before recording. This was done because we aimed to simulate a real-life scenario that applies both in the lab and in remote online data collection: in speech production studies, participants are regularly asked to view documents that present text for reading or images to describe etc, and thus it is not possible to stop all processes on a device other than the recording application. For the same reasons, participants were not asked to use an external microphone for either Phone or Zoom, as this is equipment that may not always be available. Thus, the present study provides an acoustic analysis of data acquired using the simplest application settings and readily available equipment.

### E. Measurements and Statistical Analysis

The recordings provided us with a corpus of 504 tokens [7 participants × 8 vowels × 3 repetitions × 3 devices]. These vowel tokens were manually segmented in Praat (Boersma and Weenink, 2019). F0, F1, F2, and F3 were extracted using both VoiceSauce (Shue, 2010) and Praat. Praat was chosen because it is commonly used for data extraction in speech analysis. VoiceSauce is an alternative tool which implements different algorithms and has a finer step for data extraction (see below). For F0, the range for extraction was 40-500 Hz in both VoiceSauce and Praat. For vowel formants, default VoiceSauce settings were used (covariance method, pre-emphasis of 0.96). In Praat, the maximum number of formants was set at five, and the formant extraction ranges were specified for males as 0-5000 Hz and for females as 0-5500 Hz. Mean F0, F1, F2, and F3 of all 3 tokens per vowel were calculated. For the VoiceSauce-extracted data, token means were calculated from values extracted every 1 ms throughout each token with a moving window length of 25 ms. In Praat, means were extracted with the built-in "Get mean" function.

Linear mixed effect models (Bates et al., 2015) were built in R (R Core Team, 2020) to investigate how much variation in the dependent variables (F0, F1, F2, and F3) can be ascribed to the recording devices. Data extracted using VoiceSauce and Praat were analysed separately. For each dependent variable, full models were constructed with a fixed effect of DEVICE (H6, Phone, Zoom). SPEAKER (seven speakers), VOWEL (eight vowels), PHONE_EQUIPMENT (phone models used for phone recording), and COMPUTER_EQUIPMENT (computer models used for Zoom recording) were treated as random intercepts, accounting for interspeaker differences and the use of different phone and computer models (see Table 1). Random slopes for DEVICE, PHONE_EQUIPMENT, and COMPUTER_EQUIPMENT were also fitted for SPEAKER in the full model. The random slopes and intercepts were reduced when the full models failed to converge or resulted in singular fit. Final models were the same for VoiceSauce- and Praat-extracted data, and are reported together with the results in Tables II-V. These tables present estimated difference (*estimate*), standard error (*SE*), degrees of freedom (*df*), t value (*t*), which reflects how extreme the observed difference is relative to the intercept, and *p*-value (significance) from the t-test (*Pr(>|t|)*).

Illustrative boxplots, separately for data extracted using Praat and VoiceSauce, can be found in Fig. 1. The boxplots show *differences* between devices, calculated by subtracting H6 values from the Phone values and the Zoom values for the same token across the three devices. In other words, each value plotted represents the difference between matching paired tokens from the devices.

## III. RESULTS

Statistical models and results are shown in Tables II-V. Note that for all analyses, COMPUTER_EQUIPMENT and PHONE_EQUIPMENT were not retained in the final models due to singular fits (see section II E). Fig. 1a-d illustrates the difference between devices for F0, F1, F2, and F3 in the data extracted using VoiceSauce (left) and Praat (right).

Regarding F0, there was no statistically significant effect of DEVICE (see Table II) for data extracted using either VoiceSauce or Praat. Nevertheless, the data contain some outliers. For instance, the positive outliers for the F0 of /e/ in the VoiceSauce-extracted data (circled in Fig. 1a) were both from the same repetition simultaneously recorded by the three devices. Such outliers suggest that on occasion both devices failed to capture F0 accurately in a way that could be extracted successfully by VoiceSauce.

For F1, the effect of DEVICE was significant (see Table III). While Phone recordings did not present any significant differences from the H6 recordings, Zoom recordings had significantly lower F1 than H6 both for VoiceSauce- and Praat-extracted data.

The effect of DEVICE was also significant for F2 (see Tables IV). The F2 of VoiceSauce-extracted Phone data was significantly lower than that of H6. However, the F2 difference between H6 and Praat-extracted Phone data was not statistically significant. For Zoom, both VoiceSauce- and Praat-extracted values were significantly lower than those of H6. Fig. 1c shows that the F2 of front vowels is most affected in both VoiceSauce- and Praat-extracted data.

For F3 (see Table V), neither Phone nor Zoom showed statistical differences from H6 in the VoiceSauce-extracted data. However, in the Praat-extracted data, Zoom F3 values were lower than H6. Fig. 1d (right) reflects that Praat had difficulty in extracting F3 data from Zoom recordings across all vowels.

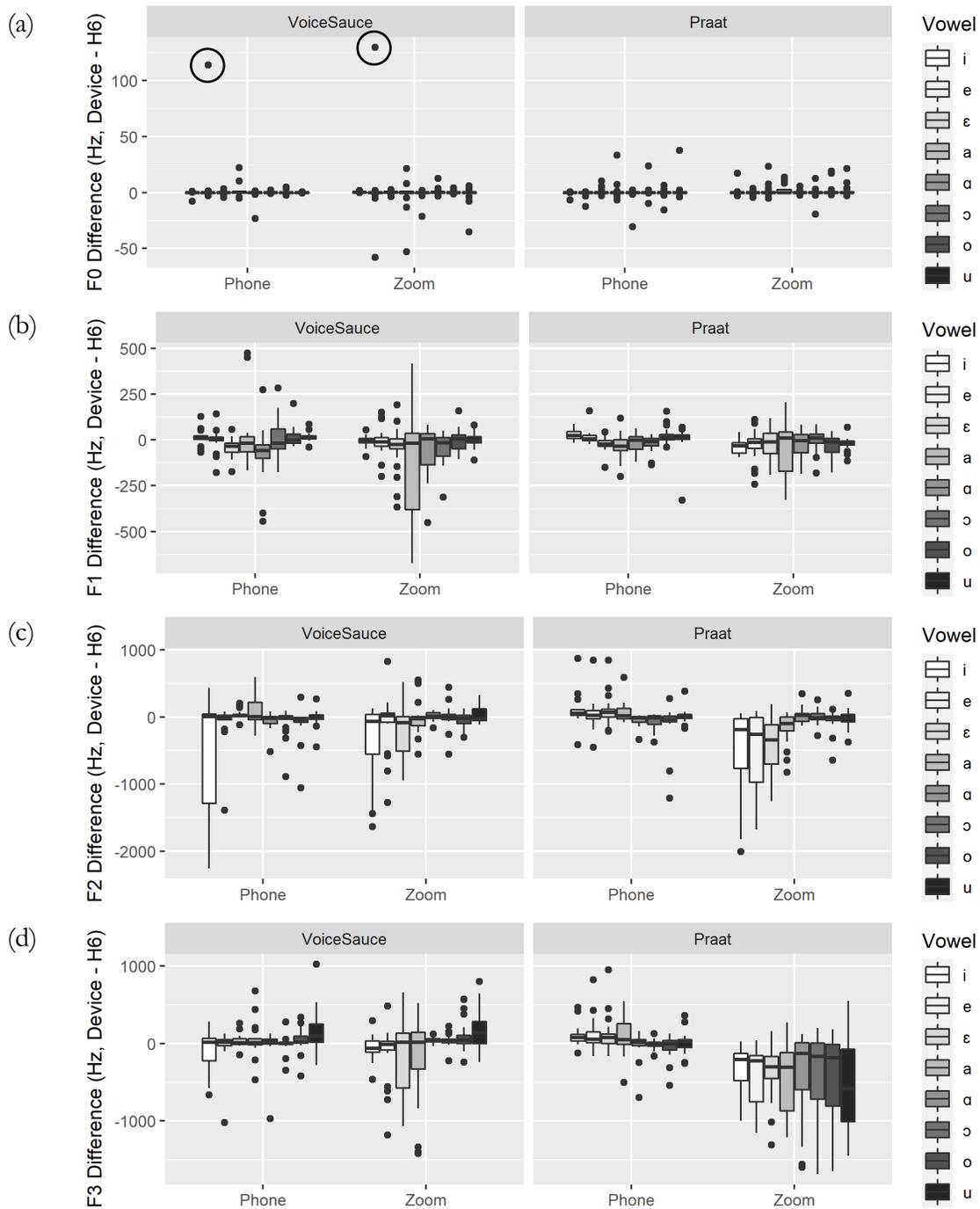

FIG. 1. Boxplots of differences in frequency between H6 and Phone and H6 and Zoom for F0 (panel a), F1 (panel b), F2 (panel c) and F3 (panel d) for VoiceSauce-extracted data (left) and Praat-extracted data (right). The middle line represents the median, upper and lower edges of the box represent the first and third quartiles, and the whiskers indicate the range, up to 1.5 times the inter-quartile range away from the median.

TABLE II. Results from the final statistical models for F0 (intercept: H6); formula: F0 ~ device + (1 | speaker) + (1 | vowel).

|  |  | Estimate | SE | df | t | Pr(>|t|) |
|---|---|---|---|---|---|---|
| **VoiceSauce** | (Intercept) | 169.94 | 18.39 | 7.13 | 9.24 | <0.001 |
|  | devicePhone | 0.6 | 1.38 | 490.01 | 0.43 | 0.664 |
|  | deviceZoom | -0.04 | 1.38 | 490.01 | -0.03 | 0.979 |
| **Praat** | (Intercept) | 170.35 | 19.23 | 7.14 | 8.86 | <0.001 |
|  | devicePhone | 0.18 | 1.13 | 490 | 0.16 | 0.872 |
|  | deviceZoom | 1.14 | 1.13 | 490 | 1.01 | 0.315 |

TABLE III. Results from the final statistical models for F1 (intercept: H6); formula: F1 ~ device + (1 | speaker) + (1 | vowel).

|  |  | Estimate | SE | df | t | Pr(>|t|) |
|---|---|---|---|---|---|---|
| **VoiceSauce** | (Intercept) | 518.63 | 60.67 | 10.58 | 8.55 | <0.001 |
|  | devicePhone | -7.68 | 11.42 | 489.99 | -0.67 | 0.502 |
|  | **deviceZoom** | **-36.95** | **11.42** | **489.99** | **-3.23** | **0.001** |
| **Praat** | (Intercept) | 545.88 | 60.1 | 10.94 | 9.08 | <0.001 |
|  | devicePhone | -2.92 | 8.12 | 489.99 | -0.36 | 0.719 |
|  | **deviceZoom** | **-31.12** | **8.12** | **489.99** | **-3.83** | **<0.001** |

TABLE IV. Results from the final statistical models for F2 (intercept: H6); formula: F2 ~ device + (1 | speaker) + (1 | vowel).

|  |  | Estimate | SE | df | t | Pr(>|t|) |
|---|---|---|---|---|---|---|
| **VoiceSauce** | (Intercept) | 1470.34 | 209.71 | 8.89 | 7.01 | <0.001 |
|  | **devicePhone** | **-90.26** | **36.62** | **489.99** | **-2.46** | **0.014** |
|  | **deviceZoom** | **-90.28** | **36.62** | **489.99** | **-2.47** | **0.014** |
| **Praat** | (Intercept) | 1453.67 | 210.74 | 8.64 | 6.9 | <0.001 |
|  | devicePhone | 12.93 | 28.5 | 489.99 | 0.45 | 0.65 |
|  | **deviceZoom** | **-195.86** | **28.5** | **489.99** | **-6.87** | **<0.001** |

TABLE V. Results from the final statistical models for F3 (intercept: H6); formula: F3 ~ device + (1 | speaker) + (1 | vowel).

|  |  | Estimate | SE | df | t | Pr(>|t|) |
|---|---|---|---|---|---|---|
| **VoiceSauce** | (Intercept) | 2894.6 | 82.03 | 14.5 | 35.29 | <0.001 |
|  | devicePhone | 14.56 | 26.32 | 489.97 | 0.55 | 0.581 |
|  | deviceZoom | -28.51 | 26.32 | 489.97 | -1.08 | 0.279 |
| **Praat** | (Intercept) | 2870.76 | 102.62 | 14.1 | 27.98 | <0.001 |
|  | devicePhone | 46.41 | 34.4 | 489.97 | 1.35 | 0.178 |
|  | **deviceZoom** | **-362.83** | **34.4** | **489.97** | **-10.55** | **<0.001** |

## IV. Additional issues

### A. AVR missing samples

Audio files recorded by the AVR application using Android devices produced a warning when opened in Praat: "File too small (1-channel 16-bit). Missing samples were set to zero." However, there were no audible glitches and files could be opened. Sample dropping in these files was investigated to understand its possible effects on measurement extraction.

A small number of zero sequences were found in the recordings, confirming AVR was dropping samples. This occurred across a range of smartphones, when using different recording options (i.e., sample rate and bitrate). To address this issue, we first investigated if sample dropping was due to phones running other applications in conjunction with AVR but this turned out not to be the case: samples were dropped whether all other applications were disabled or other applications were running at the same time as AVR. Following this finding, we proceeded with the analysis of sample dropping when both AVR and other applications were running. Analysis showed that the vast majority of zero sequences found within the recordings consisted of only 2 samples, while none exceeded 20 samples. Sequences of more than 20 zero samples were found only at the very beginning of recordings and had a maximum of 150 zeros (= 4.7 ms). Considering that the sampling rate was 44.1 kHz, these dropped samples formed a minute fraction of the duration of each recording and thus are unlikely to pose problems for analysis.

In order to completely rule out the possibility that these inconsistencies can negatively affect acoustic measurements, a simulation was run. Audio files containing artificial vowels with a duration of over 1 s were created using Praat's VowelEditor. These were compared to artificially corrupted versions of the same files, such that the latter included sequences of up to 20 zero samples. Over 4,800 such pairs were generated in Praat using ten different vowels with a variety of F0 slopes. Measurements of intensity, F0, F1 and F2 in both versions showed correlations above 0.99. This suggests that the missing samples in recordings from the AVR application do not present an issue in extracting these acoustic measures.

### B. Zoom intensity drop

In Zoom recordings, intensity was not reliably tracked, at least with the default setting with noise-cancelling processing. Periods of extremely reduced intensity occurred at random, as shown in Fig. 2. Further investigation is needed into the effects over more varied speech data. In our view, such random, extreme errors make Zoom unsuitable for phonetic research, at least in relation to any intensity-related measurements. However, in the more recent versions of Zoom, an extra setting of "enable original sound" is present, which may have the potential of recording audio with higher fidelity. Note that Sanker et al. (2021) found no significant difference in using the "enable original sound" setting and the default setting. Further examination is still needed for investigating the recording quality of audio recorded with the new option.

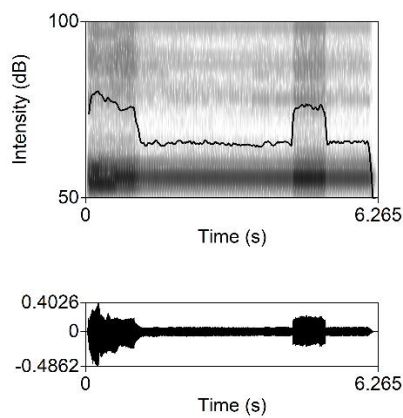

FIG. 2. One repetition of vowel [o] from PF3's Zoom recording; spectrogram with intensity curve at top, waveform, below.

## V. DISCUSSION AND CONCLUSIONS

In summary, Phone and Zoom recordings produced similar F0 values to the H6, a result consistent with previous studies which also showed that F0 is robust to lossy compression and unaffected by device choice (cf. Grillo et al., 2016; Uloza et al., 2015; Vogel et al., 2014).

Formant tracking presented some issues for the test devices, and these differed by extraction method. For the Zoom recordings, Praat-extracted data showed differences for all three formants relative to H6. For the Phone recordings, however, there were no differences. VoiceSauce-extracted data, on the other hand, showed differences in F1 and F2 values of Zoom recordings, as well as F2 values of Phone recordings. The differences between these extraction methods could be led by the different formant range settings in Praat and VoiceSauce: in Praat, the formant ranges can be set differently by gender, while formant ranges cannot be changed in VoiceSauce. These results also showed that Praat may not be able to track F2 and F3 reliably for Zoom recordings. This poses serious issues when using Zoom to record and Praat for data extraction if formant frequencies are measured. Similarly, the intensity drops observed in the Zoom recordings, while not statistically modelled in this paper, could pose serious issues for intensity analysis.

Close inspection of the data illustrated in Fig. 1 strongly indicates that there are inconsistencies in formant tracking for individual vowels. Discrepancies affected both Zoom and Phone, but for the former, there were more problems and they were of greater magnitude; consider, in Fig. 1 e.g., the F1 of [a] and [ɑ], F2 of [i], [e] (for Praat), and [ɛ], and F3 for all Praat-extracted data. The effects of recording device on F1 and F2 can vary considerably by participant, as illustrated in Fig. 3, which depicts the vowel space of each participant by DEVICE. Fig. 3 illustrates the unpredictable nature of the values recorded by the devices and the distortions they can bring. Based on these finding, we concur with De Decker and Nycz (2011) that researchers should not use different devices (e.g., Zoom and Phone) to record data for the same study, nor should they compare data obtained using different devices or extracted using different extraction methods. Finally, we note that overall, more tracking errors occurred with the female data (PF1-4) than the male data (PM1-3), across all devices. This is in line with previous reports, such as De Decker and Nycz (2011). However, we also note that not all patterns can be explained by speaker sex. For example, the vowel spaces of PF3 are similar across devices for both VoiceSauce- and Praat-extracted data, while those of PM1 show substantial differences.

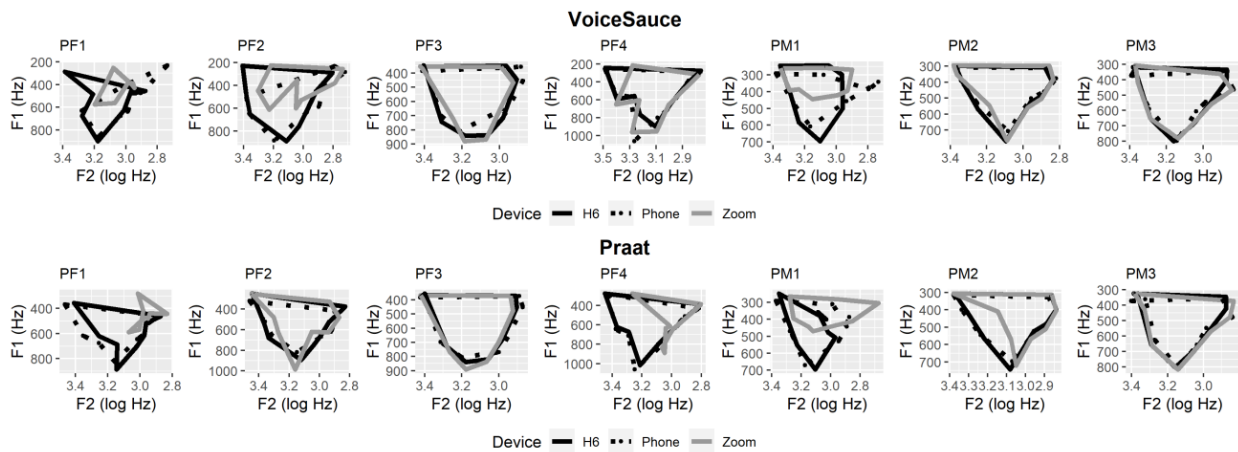

FIG. 3: Vowel space by device for each speaker, for VoiceSauce-extracted data (top) and Praat-extracted data (bottom).

While this study serves as a starting point to compare the differences between recording devices and provides researchers with some insight about remote data collection methods, the quality differences between recording methods is a complex, multifaceted issue that requires further investigation. While we tested whether running other applications caused sample dropping in Phone recordings using AVR (it did not), there are further questions that could be addressed regarding specific recording conditions. For instance, whether Zoom audio quality could be improved by closing other applications, clearing device memory, or using a headset microphone. Further tests of what causes the unreliability of Zoom recordings could include investigating the effects of compression, and post-processing of the data files. Future studies are also necessary to look at running speech

In conclusion, our findings indicate that lossless recordings from phones can be a viable method for recording vowel data for acoustic analysis, at least with respect to F0, F1 and F2. On the other hand, caution is needed if conditions limit a researcher's choice to the use of lossy Zoom recordings, as these can lead to erratic outcomes.

**ACKNOWLEDGEMENTS**

This research was conducted with support from grant ERC-ADG-835263 titled "Speech Prosody in Interaction: The form and function of intonation in human communication" awarded to Amalia Arvaniti.This research was conducted with support from grant ERC-ADG-835263 titled "Speech Prosody in Interaction: The form and function of intonation in human communication" awarded to Amalia Arvaniti.

TABLE I. Recording equipment and application information for each participant.

| Participant ID | Mobile phone | App | Zoom | Recorder | Microphone |
|---|---|---|---|---|---|
| **PF1** | Samsung Note10 | AVR[a] | MS Surface Pro 6 | Zoom H6 | Sennheiser HSP2 |
| **PF2** | Samsung Galaxy S10e | AVR | Dell Precision 5520 | Zoom H6 | Rode NT3 cardioid mic (on stand) |
| **PF3** | Samsung Note10 | AVR | MS Surface Pro 6 | Zoom H6 | Sennheiser HSP2 |
| **PF4** | Google Pixel 3a | AVR | Lenovo Thinkpad T495 | Zoom H6 | Sennheiser HSP2 |
| **PM1** | Samsung Note9 | AVR | MS Surface Pro 6 | Zoom H6 | Sennheiser HSP2 |
| **PM2** | Apple iPhone 5s | AVR | Lenovo Thinkpad T495 | Zoom H6 | Sennheiser HSP2 |
| **PM3** | bq AQUARIS E4.5 Ubuntu Edition | Recorder[b] | Lenovo Thinkpad T495 | Zoom H6 | Sennheiser ME64 & K6P |

[a] Awesome Voice Recorder (Newkline, 2020)

[b] Recorder (DawnDIY, 2016) is an app available on Linux phones; see section II D. for details.

TABLE II. Results from the final statistical models for F0 (intercept: H6); formula: F0 ~ device + (1 | speaker) + (1 | vowel).

|  |  | Estimate | SE | df | t | Pr(>|t|) |
|---|---|---:|---:|---:|---:|---:|
| **VoiceSauce** | (Intercept) | 169.94 | 18.39 | 7.13 | 9.24 | <0.001 |
|  | devicePhone | 0.6 | 1.38 | 490.01 | 0.43 | 0.664 |
|  | deviceZoom | -0.04 | 1.38 | 490.01 | -0.03 | 0.979 |
| **Praat** | (Intercept) | 170.35 | 19.23 | 7.14 | 8.86 | <0.001 |
|  | devicePhone | 0.18 | 1.13 | 490 | 0.16 | 0.872 |
|  | deviceZoom | 1.14 | 1.13 | 490 | 1.01 | 0.315 |

TABLE III. Results from the final statistical models for F1 (intercept: H6); formula: F1 ~ device + (1 | speaker) + (1 | vowel).

|  |  | Estimate | SE | df | t | Pr(>|t|) |
|---|---|---|---|---|---|---|
| **VoiceSauce** | (Intercept) | 518.63 | 60.67 | 10.58 | 8.55 | <0.001 |
|  | devicePhone | -7.68 | 11.42 | 489.99 | -0.67 | 0.502 |
|  | **deviceZoom** | **-36.95** | **11.42** | **489.99** | **-3.23** | **0.001** |
| **Praat** | (Intercept) | 545.88 | 60.1 | 10.94 | 9.08 | <0.001 |
|  | devicePhone | -2.92 | 8.12 | 489.99 | -0.36 | 0.719 |
|  | **deviceZoom** | **-31.12** | **8.12** | **489.99** | **-3.83** | **<0.001** |

TABLE IV. Results from the final statistical models for F2 (intercept: H6); formula: F2 ~ device + (1 | speaker) + (1 | vowel).

|  |  | Estimate | SE | df | t | Pr(>|t|) |
|---|---|---|---|---|---|---|
|  | (Intercept) | 1470.34 | 209.71 | 8.89 | 7.01 | <0.001 |
| **VoiceSauce** | **devicePhone** | **-90.26** | **36.62** | **489.99** | **-2.46** | **0.014** |
|  | **deviceZoom** | **-90.28** | **36.62** | **489.99** | **-2.47** | **0.014** |
|  | (Intercept) | 1453.67 | 210.74 | 8.64 | 6.9 | <0.001 |
| **Praat** | devicePhone | 12.93 | 28.5 | 489.99 | 0.45 | 0.65 |
|  | **deviceZoom** | **-195.86** | **28.5** | **489.99** | **-6.87** | **<0.001** |

TABLE V. Results from the final statistical models for F3 (intercept: H6); formula: F3 ~ device + (1 | speaker) + (1 | vowel).

|  |  | Estimate | SE | df | t | Pr(>|t|) |
|---|---|---|---|---|---|---|
| **VoiceSauce** | (Intercept) | 2894.6 | 82.03 | 14.5 | 35.29 | <0.001 |
|  | devicePhone | 14.56 | 26.32 | 489.97 | 0.55 | 0.581 |
|  | deviceZoom | -28.51 | 26.32 | 489.97 | -1.08 | 0.279 |
| **Praat** | (Intercept) | 2870.76 | 102.62 | 14.1 | 27.98 | <0.001 |
|  | devicePhone | 46.41 | 34.4 | 489.97 | 1.35 | 0.178 |
|  | **deviceZoom** | **-362.83** | **34.4** | **489.97** | **-10.55** | **<0.001** |

Collected figure captions:

FIG. 1. Boxplots of differences in frequency between H6 and Phone and H6 and Zoom for F0 (panel a), F1 (panel b), F2 (panel c) and F3 (panel d) for VoiceSauce-extracted data (left) and Praat-extracted data (right). The middle line represents the median, upper and lower edges of the box represent the first and third quartiles, and the whiskers indicate the range, up to 1.5 times the inter-quartile range away from the median.

FIG. 2. One repetition of vowel [o] from PF3's Zoom recording; spectrogram with intensity curve at top, waveform, below.

FIG. 3: Vowel space by device for each speaker, for VoiceSauce-extracted data (top) and Praat-extracted data (bottom).